# On the Dynamical Origin of Bias in Clusters of Galaxies

S. Colafrancesco[1]

Osservatorio Astronomico di Roma, Via dell'Osservatorio 5

I-00040 Monteporzio (Roma), Italy

E-mail: cola@astrmp.astro.it, 29575::cola

V. Antonuccio-Delogu[1,2]

Osservatorio Astrofisico and CNR-GNA, Unitá di Ricerca di Catania,

Viale A. Doria 6, I-95125 Catania, ITALY

E-mail: vantonuccio@astrct.ct.astro.it, 39003::vantonuccio

and

A. Del Popolo[1]

Dipartimento di Astronomia, Università di Catania, ITALY

Viale A. Doria 6, I-95125 Catania, ITALY

E-mail: antonino@astrct.ct.astro.it, 39003::antonino

---

[1]Member of the European Cosmology Network

[2]Theoretical Astrophysics Center, University of Copenhagen, DENMARK

# ABSTRACT

We study the effect of the dynamical friction induced by the presence of substructure on the statistics of the collapse of density peaks. Applying the results of a former paper (Antonuccio-Delogu and Colafrancesco, 1994) we show that within high density environments, like rich clusters of galaxies, the collapse of the low $\nu$ peaks is strongly delayed until very late epochs. A bias of dynamical nature thus naturally arises because high density peaks preferentially collapse to form halos within which visible objects eventualy will form. We then derive a *selection function* for these collapsing structures: with this in hands, we can calculate the values of the bias coefficient on cluster scales for any hierarchical clustering scenario. For a standard CDM model we show here that the *dynamical bias* that we derive can account for a substantial part of the total bias required by observations on cluster scales.

*Subject headings:* cosmology: theory — galaxies: clustering — galaxies: formation

## 1. Introduction

Several observational evidences indicate that the distribution of groups and clusters of galaxies has to be biased with respect to the underlying distribution of cosmic material on large scales. A widespread theoretical motivation for this is to assume that the formation of actual structures proceeds around the high peaks of the initial density field (Kaiser 1984, Peacock & Heavens 1985, Bardeen et al. 1986 - hereafter BBKS).
Within this frame (and assuming a Gaussian statistics) the peak correlation $\xi_{pk}(r) \approx b^2 \xi_\rho(r)$ is directly related to the structure of the underlying density field: hence, the power and mass spectra are proportional through the biasing factor $b$: $\langle \delta^2(k) \rangle = b^2 \langle \Delta^2_{mass}(k) \rangle$. This Lagrangian biasing is spatially uniform; it intervenes also in the normalization of the overall amplitude of the linear perturbation spectrum through the function $J_3(r) = \int dr \xi(r) r^2$. In terms of this Lagrangian biasing, the 2-point correlation function for rich clusters indicates $b \sim 3$ (see, e.g., BBKS), a result also confirmed by the APM survey (Maddox et al. 1990). Moreover, the observed mass distribution of Abell clusters (Bahcall & Cen 1992) requires a strong suppression of galaxy formation. Within the spherical model for the evolution of density perturbations (see Peebles 1980) this is consistent with a value $b \approx 1.7 \div 1.9$ (see also Antonuccio-Delogu and Colafrancesco, 1994). An extensive analysis of the X-ray luminosity functions and of the X-ray counts for galaxy clusters (see Colafrancesco and Vittorio 1994, hereafter CV) indicates that the local and distant distributions of X-ray clusters depend on the combination $b \cdot \delta_c$, where $\delta_c$ is the linearly extrapolated threshold for collapse. Values $b \cdot \delta_c \approx 3$ obtain for a wide range of cosmological models. This result is also confirmed by recent hydrodynamic simulations (Cen & Ostriker 1993, Bryan et al. 1994) showing that a COBE normalized (unbiased) standard CDM scenario predicts a space density of X-ray clusters in excess by a factor $\sim 5$ with respect to the observed value.

From the result of CV, it appears that the cluster abundance depends on the effects of non-linear dynamics during cluster formation (for which $\delta_c$ is an indicative parameter) and/or from the uncertainty in the normalization of the linear power spectrum, as contained in the value of the bias parameter $b$. Both these issues are emanations of a biasing process already effective.

The physical origin of such a biasing is not yet clear even though several mechanisms have been proposed (see Rees 1985, Dekel & Rees 1987, Dekel & Silk 1986, Carlberg 1991, Cen & Ostriker 1992, Bower et al. 1993, Silk & Wyse 1993). But, if efficient, these mechanisms will generate a physical selection of those peaks in the initial density field that eventually will give rise to the observed cosmic structures.

In this paper we will derive (in §2) a *selection function* for the peaks that give rise to the protostructures: it results from a proper treatment of the dynamical effects of small scale substructure during the early stages of the collapse. In §3 we will derive values of the biasing parameter on the relevant filtering scales as produced by the previous non-linear mechanisms and we will discuss their cosmological relevance in the conclusions (§4).

## 2. Selection function

Following BBKS we define a selection function $t(\nu)$ as the probability that a peak of the filtered density field will eventually form an object:

$$n_{obj} = \int_0^\infty t(\nu) N_{pk}(\nu) \tag{1}$$

where $n_{obj}, N_{pk}(\nu)$ are the objects' and peaks' densities, respectively (we restrict ourselves to overdense regions, so the lower limit of integration in eq. 1 is put equal to 0). The specification of the selection function is crucial to relate the properties of the "objects" forming out of the density peaks. BBKS adopted an empirical selection function of the form: $t(\nu) = (\nu/\nu_t)^q / [1 + (\nu/\nu_t)^q]$, which depends on two parameters, namely the *threshold* $\nu_t$ and a shape exponent $q$. The results they obtain concerning the statistical properties of clusters and galaxies depend then on these parameters, because the selection function describes the statistical relationship among peaks and objects.

We will now give a more physically motivated prescription to obtain a function $t(\nu)$ selecting precisely the "objects" we defined as galaxy clusters. Let us consider the average density profile around a peak of a given central height $\nu$ (see Figure 1) in the density field smoothed on an arbitrary scale $R_f$. The outermost shell defining our objects will be at a distance $r_{M_T}$ defined in such a way that the total mass contained within the average density profile of the peak is fixed to an arbitrary value $M_T$. This distance depends also on $\nu$ and on $R_f$, and thus in the following we will compute the bias coefficients for different values of $M_T$ and $R_f$. The average overdensity at the distance $r_{M_T}$ can be either larger or smaller than the critical overdensity for collapse. In the latter case, if the density profile of our peak coincides with the average one, the shell will not collapse. However, the previous argument holds for the average profile: at any distance there is a finite probability that the peak overdensity will be different from the average one, and for a Gaussian density field this probability is still a Gaussian:

$$p\left(\bar{\delta}, \langle\bar{\delta}\rangle(r)\right) = \frac{1}{\sqrt{2\pi}\sigma_{\bar{\delta}}} \exp\left(-\frac{|\bar{\delta} - \langle\bar{\delta}\rangle(r)|^2}{2\sigma_{\bar{\delta}}^2}\right) \quad (2)$$

where $\langle\bar{\delta}\rangle(r)$ is the average overdensity and the dispersion $\sigma_{\bar{\delta}}(r)$ is given, e.g, by eq. 3.8 of Lilje & Lahav 1991. Given a critical overdensity threshold for collapse $\delta_c$, for any given density peak of specified height $\nu$ we can define a probability that the outermost shell will collapse as:

$$t(\nu) = \int_{\delta_c}^{\infty} d\rho \ p\left[\bar{\delta}, \langle\bar{\delta}\rangle(r_{M_T})\right] \quad (3)$$

The integrand in eq.(3) is evaluated at a fixed distance $r_{M_T}$. We define $t(\nu)$ in eq.(3) as our *selection function*. Its dependence on $\nu$ and $R_f$ is implicit through the dependence from the average density profile $\bar{\delta}(r)$. For $\nu \to \infty$ the average density raises much over $\delta_c$, so one has $\delta_c \ll \bar{\delta}(r_{M_T})$, and $t(\nu) \to 1$. In the opposite case, when $\nu \to 0$, the variance $\sigma_{\bar{\delta}}^2$ in eq. 2 is always very large over the integration domain, so $t(\nu) \to 0$. In Figure 2 we plot $t(\nu)$ for different values of the lower integration limit in eq. 3.

As we have shown in a former paper (Antonuccio-Delogu and Colafrancesco, 1994, hereafter AC94), the presence of small-scale substructure modifies the value of the critical threshold for collapse $\delta_c$ from the value $\delta_{c0} \approx 1.68$ holding for the ideal case of a collapse without substructure. The dynamical friction induced on objects contained within infalling shells by the small scale substructure induces a drag force which act to contrast the gravitational force. As a consequence, the value of $\delta_c$ is larger w.r.t. the ideal case (AC94):

$$\delta_c = \delta_{c0} \left[ 1 + \frac{\lambda_0}{1 - \mu(\delta_c)} \right] \qquad (4)$$

where the dynamical friction coefficient is given by (see AC94, eq. 23):

$$\lambda_0 = 4.44 \frac{G^2 \langle m \rangle_{av}^2 \langle n \rangle_{av}}{\langle v \rangle_{av}^{3/2}} \cdot \log \left[ 1.12 \frac{\langle v \rangle_{av}^2}{G \langle m \rangle_{av} \langle n \rangle_{av}^{1/3}} \right] \cdot$$

$$\pi \left( \frac{3}{8\pi G \rho_c} \right)^{1/2} \qquad (5)$$

and the function $\mu(\bar{\delta})$ is given by:

$$\mu(\bar{\delta}) = \frac{\sqrt{2}\pi}{3c(\bar{\delta})} \left( \frac{1}{\bar{\delta}} + 1 \right)^{3/2} \qquad (6)$$

In eq. 5 the quantities in bracket are averages over the population of "subpeaks" generating the small scale substructure. The quantity $c(\bar{\delta})$ was defined in AC94.

Eq. 4 is an implicit equation for $\delta_c$ which can be solved once $\lambda_0$ has been specified. This latter quantity depends on the mass spectrum and temporal evolution of the small scale substructure. These two important inputs are today poorly known both from theoretical and from observational grounds. We then decided to adopt the same strategy as in AC94, namely to include only the contributions from the peaks selected up to a maximum mass of $\approx 7.34 \times 10^8 M_\odot$, which is more than three orders of magnitude below the filtering mass of the considered objects (hereafter, we will refer to this population of density peaks as the *small-scale substructure*, *SSS*). The values of $\lambda_0$ so obtained is then a lower limit, also because we do not take into account in eq. 5 any term depending on the correlation function (Antonuccio-Delogu, 1992;Antonuccio-Delogu and Atrio-Barandela, 1992). However, we take into account the fact that the peaks' number density will be enhanced within protoclusters w.r.t the background. Following BBKS the enhancement factor is given by:

$$f_{enh} = \exp\left[\alpha\nu_{cl}\frac{\sigma_{cl}}{\sigma_{obj}} - \frac{\beta}{2}\nu_{cl}^2\left(\frac{\sigma_{cl}}{\sigma_{obj}}\right)^2\right] \tag{7}$$

where a subscript "*cl*" denotes quantities computed within the protocluster. In the following of this *Letter* we will show results obtained for $\nu_{cl} = 2$ and $R_{cl} = 5h^{-1}$ Mpc, typical of a rich Abell cluster.

### 3. Bias coefficients

The bias coefficient of a given class of objects can be given by (see BBKS):

$$b_{ss}(R_f) = \frac{\langle\tilde{\nu}\rangle}{\sigma_0} + 1 \tag{8}$$

where the average $\langle\tilde{\nu}\rangle$ is given by (Bardeen et al. 1986, eq. 6.45):

$$\langle\tilde{\nu}\rangle = \int_0^\infty d\nu\left[\nu - \frac{\gamma\theta}{1-\gamma^2}t(\nu)N_{pk}(\nu)\right] \tag{9}$$

and $\gamma$ and $\theta$ denote the usual spectral quantities (see BBKS and AC94). Once we have specified a given spectrum we have all the tools to evaluate the integral in eq. 9 and then to calculate the bias parameter $b_{ss}$.

For the sake of brevity, we will now restrict ourselves to the standard CDM model leaving to a more general paper the applications to other scenarios of structure formation (Antonuccio-Delogu, Colafrancesco and Vittorio 1994). At present, there is a rather large uncertainty about the value of the normalization constant of the standard CDM spectrum, with values ranging from $b_8 \approx 1$ (coming from the analysis of COBE data, Scaramella and Vittorio 1993) up to $b_8 \approx 1.8$. Here we present our results for four different values of $b_8$, namely $b_8 = 1.2, 1.4, 1.6, 1.8$, and for three values of $R_f$, the Gaussian filtering radius of the objects: $R_f = 200, 356$ and $500$ $h^{-1}$ kpc. For the first two values of $R_f$ we choose $r_{M_T}$ in such a way that the total mass enclosed by the outermost shell defining the object is $\approx 2 \times 10^{12} M_\odot$, a value typical of a bright spiral galaxy (see Taable 1). So from a comparison of the results for these two cases we can have insights on how the bias induced by small scale substructure is affected by the compactness of the structure. The results are shown in Table 1 and Figure 3. In the fifth column of Table 1 we report the values of $b_{ss}$, the dynamical friction-induced bias coefficient, computed from eq. 8 and from the proper values of $\delta_c$ (see eq 4). For larger $R_f$ (at fixed $M_T$) the value of $b_{ss}$ increases: this is a consequence of the behaviour of the selection function $t(\nu)$ in these two cases. In fact, $t(\nu)$ acts as a filter in eq. 8, and from Figure 2 we see that the approximate value of $\nu$ at which $t(\nu) \sim 1$ increases with increasing $R_f$. The physical reason for this is that, leaving constant the value of $M_T$, a peak selected with a larger filter radius is shallower than a peak selected with a smaller radius, and the gravitational force on the outer shells will be smaller.

The error bars in Figure 3 were computed by taking into account the theoretical uncertainties in the calculation of $\lambda_0$. These originate from the fact the quantities $\langle m \rangle_{av}, \langle n \rangle_{av}$ depend on very uncertain functions like the typical mass associated with a peak of a given height, $m_{peak}(\nu)$. We have adopted the formula given by Peacock and Heavens (1985), but in order to compute the error bars we have also allowed the value of $m_{peak}(\nu)$ to be estimated by a top-hat filtering mass. Analogously, we have taken into account the $\pm 1\sigma$ on the number density of the peaks (BBKS).

We also notice that for smaller peaks ($R_f = 200 h^{-1}$ Kpc, $M_T = 1.4 \times 10^{11} M_\odot$) the bias coefficient is also smaller, consistently with the interpretation we have just given. In general, the values of $b_{ss}$ are smaller than the global bias coefficient $b_8$, as it should be. However, they can account for a large fraction of the total biasing level, particularly at low values of $b_8$ which seem to be favoured by COBE.

## 4. Conclusions

In this Letter we have looked at one of the consequences of the delay of the collapse of protostructures induced by the presence of small-scale, fine-grained substructure. This collapse delay requires that the critical overdensity for the collapse of a spherical perturbation has to be larger than the canonical value $\delta_c \geq \delta_{co} \approx 1.68$ (see also Antonuccio-Delogu and Colafrancesco, 1994).
As a consequence of this result, within overdense regions on the cluster scales, the progenitors of the structures which collapse and virialise by the present time are among the peaks which are selected according to the selection function $t(\nu)$ introduced in section 3. This statistical selection function takes into account the *dynamical biasing* of realistic protostructures out of the structure of the density field at early times. Thus it is not a pure Heaviside function as in the empirical case treated by BBKS: its shape depends on the assumed overdensity threshold for collapse $\delta_c$ and on the probability distribution (see eq.2) for having a collapsing shell around a density peak. Moreover this function selects peaks of larger height for increasing filtering scale $R_f$.

Based on our selection function, we have also derived detailed values of the *dynamical bias*, $b_{ss}$, on the cluster scales: we found that $b_{ss} \leq b_8$. Depending on the values of $\delta_c$ and $R_f$, we found that $b_{ss}/b_8 \sim 0.5 \div 0.8$ for values $b_8 > 1.2$ as indicated by the bulk of the observations on cluster scales. This means that a sensitive fraction of the bias coefficient $b_8$ required to match the observations of the cluster abundance can be accounted for by the non-linear, dynamical effects of substructure during the formation of galaxy clusters.

Here we have also shown that the non-linear effects during cluster collapse producing large values of $\delta_c$ traslate consistently in high values of the biasing parameter $b_{ss} > 1$. This justifies theoretically the conclusions of CV, although those were based only on a linear approach.

These results are obtained specifically for a standard CDM scenario ($\Omega_o = 1$, $h = 0.5$): however, they hold in principle for any hierarchical clustering scenario. Their detailed importance in different models of structure formation will be discussed in another paper. In fact, a detailed calibration of the importance of this and other non-linear mechanisms during cluster formation and evolution is needed to assess definitely the origin of the cosmological biasing in the theory of structure formation.

**Acknowledgments**. The investigation of V.A.-D. was partly supported by Danmarks Grundforskningsfond through its support for an establishment of the Theoretical Astrophysics Center (TAC).

| $R_f$ [a] | $M_T$ [b] | $b_8$ | $b_{1.68}$ | $b_{ss}$ | $\delta_c$ |
|---|---|---|---|---|---|
| 0.2 | 0.14 | 1.8 | 1.2 | 1.27 | 3.51 |
| " | " | 1.6 | 1.17 | 1.21 | 3.1624 |
| " | " | 1.4 | 1.145 | 1.17 | 2.84 |
| " | " | 1.2 | 1.12 | 1.13 | 2.53 |
| 0.356 | 1.97165 | 1.8 | 1.3 | 1.44 | 3.51 |
| " | " | 1.6 | 1.25 | 1.3 | 3.1624 |
| " | " | 1.4 | 1.214 | 1.255 | 2.84 |
| " | " | 1.2 | 1.176 | 1.2 | 2.5 |
| 0.5 | 1.1 | 1.8 | 1.33 | 1.515 | 3.51 |
| " | " | 1.6 | 1.28 | 1.39 | 3.1624 |
| " | " | 1.4 | 1.23 | 1.29 | 2.84 |
| " | " | 1.2 | 1.19 | 1.21 | 2.5 |

Table 1: Bias coefficients.

---

[a] In $h^{-1}$ Mpc

[b] In units of $10^{12} M_\odot$

# REFERENCES

Antonuccio-Delogu, V., 1992, Ph.D. thesis, ISAS, Trieste

Antonuccio-Delogu, V. and Atrio-Barandela, 1992, ApJ, 392, 403

Antonuccio-Delogu, V. and Colafrancesco, S. 1994, ApJ, 427, 72

Antonuccio-Delogu, V., Colafrancesco, S. and Vittorio, N. 1994, in preparation

Bahcall, N. & Cen, R.Y. 1992, ApJ, 398, L81

Bardeen, J. M., Bond, J. R., Kaiser, N. and Szalay, A.S., 1986, ApJ, 304, 15

Bower, R. G., Coles,P., Frenk, C.S. and White, S.D.M., 1993, ApJ, 403, 405

Bryan, G.L., Cen, R.Y., Norman, M.L., Ostriker, J.P., and Stone, J.M. 1994, ApJ, 428, 405

Carlberg, R.G. 1991, ApJ, 367, 385

Cen, R.Y. and Ostriker, J.P. 1992, ApJ, 399, L113

Cen, R.Y. and Ostriker, J.P. 1993, ApJ, 417, 387

Colafrancesco, S. and Vittorio, N. 1994, ApJ, 422, 443

Dekel, A. and Rees, M.J. 1987, Nature, 326, 455

Dekel, A. and Silk, J. 1986, ApJ, 303, 39

Lilje, P.B. and Lahav, O., 1991, ApJ, 374, 29

Kaiser, N. 1984, ApJ, 284, L9

Maddox, S.J., Efstathiou, G., Sutherland, W.J., and Loveday, J. 1990, MNRAS, 242, 43p

Peacock, J.A. and Heavens, A.F. 1985, MNRAS, 217, 805

Rees, M.J. 1985, MNRAS, 213, 75p

Scaramella, R. and Vittorio, N., 1993, MNRAS, 263, L17

Silk, J. and Wyse, R.F.G. 1993, Physics Reports 231, 293

## Figure Captions.

**Figure 1.**–The average density profile of a peak (*continuous curve*) together with the $\pm 1\sigma_{\bar{\delta}}$ profiles (*dashed curves*). The asterisk marks the lower integration limit of the integral in eq. 3 (see section 2.).

**Figure 2.**–The selection function $t(\nu)$ for $R_f = 356$ Kpc$h^{-1}$ ($\delta_c = \delta_{c0}$ (*dots*), $\delta_c = 3.51$ (*long-dashed*)) and for $R_f = 500$ Kpc$h^{-1}$ ($\delta_c = \delta_{c0}$ (*short-dash*), $\delta_c = 3.51$ *continuous*)).

**Figure 3.**–Bias coefficients. We plot the values of $b_{ss}$ for $R_f = 500 h^{-1}$ kpc. *Filled circles*: $\delta_c = \delta_{df}$, *open circles*: $\delta_c = \delta_{c0}$. . The dotted line marks the maximum value of $b_{ss}$ (i.e. $b_{ss} = b_8$). The error bars have been computed by summing the squares of the uncertainties in $b_{ss}$ due to independent uncertainties on $\langle m \rangle_{av}, \langle n \rangle_{av}$ and on the central correlation function $\xi_0$ (Antonuccio-Delogu, 1992).